\documentclass[graybox]{svmult}
\usepackage{amssymb,amsmath,enumerate}
\usepackage{multirow}
\usepackage{bm}
\usepackage{epsfig}
\usepackage{multicol}        % used for the two-column index
\usepackage{url}
\usepackage{caption}
\usepackage{authblk}
\usepackage{mathptmx}       % selects Times Roman as basic font
\usepackage{helvet}         % selects Helvetica as sans-serif font
\usepackage{courier}        % selects Courier as typewriter font
\usepackage{type1cm}        % activate if the above 3 fonts are
\usepackage{slashbox}
\usepackage{makeidx}         % allows index generation
\usepackage{graphicx}        % standard LaTeX graphics tool
\usepackage{multicol}        % used for the two-column index
\usepackage[bottom]{footmisc}% places footnotes at page bottom

\newcommand{\Xq}{X^{(q)}}

\makeindex             % used for the subject index
                       \usepackage{amsmath,amssymb,amsfonts,graphics,graphicx,dcolumn,bm,enumerate}

\graphicspath{{./Figures/}}
\begin{document}

\title*{Dynamical evolution of anti-social phenomena: A data science approach}
\titlerunning{Dynamical evolution of anti-social phenomena}
%Use \titlerunning{Short Title} for an abbreviated version of
\author{Syed Shariq Husain and Kiran Sharma}

\institute{Syed Shariq Husain \at School of Computational and Integrative Sciences, Jawaharlal Nehru University, New Delhi-110067, India. \email{shariq.iitk@gmail.com}
\and Kiran Sharma \at School of Computational and Integrative Sciences, Jawaharlal Nehru University, New Delhi-110067, India. \email{kiransharma1187@gmail.com }
}
\maketitle

%\abstract*{In this article...}
\abstract{Human interactions can be either positive or negative, giving rise to different complex social or anti-social phenomena. The dynamics of these interactions often lead to certain spatio-temporal patterns and complex networks, which can be interesting to a wide range of researchers-- from social scientists to data scientists. 
Here, we use the publicly available data for a range of anti-social and political events like ethnic conflicts, human right violations and terrorist attacks across the globe. We aggregate these anti-social events over time, and study the temporal evolution of these events. We present here the results of several time-series analyses like recurrence intervals, Hurst R/S analysis, etc., that reveal the long memory of these time-series. Further, we filter the data country-wise, and study the time-series of these anti-social events within the individual countries.We find that the time-series of these events have interesting statistical regularities and correlations using multi-dimensional scaling technique, the countries are then grouped together in terms of the co-movements with respect to temporal growths of these anti-social events.  The data science approaches to studying these anti-social phenomena may provide a deeper understanding about their formations and spreading. The results can help in framing public policies and creating strategies that can check their spread and inhibit these anti-social phenomena.}

\vskip 0.2in

\section{Introduction}
\label{sec:Intro}
Humans prefer to form groups and act collectively. These groups have evolved from simple settlements in ancient times to huge nations in modern times;  defined by multiple causes like languages, common heritage, geographical boundaries, and even ideologies. Often human cooperation has been the motivating force behind the rapid progress of man. This cooperation \cite{Perc_2017} has extended from blood relatives to totally unrelated individuals. Contrarily, evolution has been responsible for drawing distinctions among themselves in their bids for the ``survival of the fittest''. The segregation \cite{Schelling_1969,Schelling_1971} can be seen in various forms of race, caste, class, religion, political ideology, etc. The assortment of positive and negative aspects of human social behavior makes it extremely complex and convoluted with multiple parameters playing crucial roles. Thus, it is extremely difficult to assess and model the complexity of human social behavior, ranging from bonding, co-operation, support to greed, jealousy, conflict, aggression, coup, war, etc.

Entire world has seen time and again different forms of conflicts, aggression, war, and terrorism, which have plagued mankind from antiquity. Anti-social phenomenon notably possesses very different characteristics than normal social behavior. The interactions among anti-social agents are very low and the occurrences of events tend to be independent of each other.
A conflict is an activity which takes place between conscious (not necessarily rational) beings when their interests are mutually inconsistent with each other. A conflict is usually associated with violent activities. The human society has been riddled with conflicts. The first known conflict, a case of inter-group violence, was in eastern Africa around 10,000 years ago as an attempt to seize resources - territory, women, food stored in pots,  which resulted in the killing of over two dozen prehistoric men, women, and children \cite{Lahr_2016}.  
However, as there has been more progress in civilization, humans have become more materialistic and self-centered, and gone beyond competition for tangible resources; they have adopted causes like religion, racial superiority, etc., as pretexts for killing others. 

Many people, including Karl Marx and Friedrich Engels, have proposed theories of social conflicts. 
Apart from social scientists, physicists and data scientists have recently tried to perform in-depth studies and provide mathematical models, statistical 
and time series analysis of the empirical data and tried to propose potential solutions to the menaces of terrorism, conflicts and other social phenomena, leading to the development of the field of sociophysics \cite{Castellano_2009,Chakrabarti_2006,Sen_2014,Abergel_2017}. Sociophysics is marked by the belief that large-scale statistical measurement of social variables reveals underlying relational patterns that can be explained by theories and laws found in natural sciences, and physics in particular.

In this chapter, we focus on the data dependent statistical analyses of three major anti-social phenomena, viz., ethnic conflicts (EC), human right violations (HR), and terrorism (GTD) \cite{Sharma_2017_a,Clauset_2007,Husain_2018}. 
An ethnic conflict is a conflict between two or more contending ethnic groups where each group fights for its position within the society on the basis of ethnicity, derived from common descent, culture, language and sometimes, even a common identity.
Similarly, a human right violation is said to occur when the basic fundamental rights of a person or a group of persons are infringed upon. Both these anti-social phenomena are based on the conflicts between one or more contesting parties (two sets of actors). However, terrorism differs from these conflicts in the sense that the casualties occurring in a terrorist event are direct or indirect targets of the terrorist groups (sources). The aim of terrorism is not limited to eliminating the target group or destruction of resources, rather it is specifically carried out to send out a psychological message to the adversary \cite{Richardson_2013}. In other words,   unlike in ethnic conflicts and human rights violations, the terrorist attacks are carried out to send across a message to the opponent \cite{Cutter_2014}.

Here, we use the publicly available data from: (a) GDELT database \cite{GDELT, GDELTcodebook}, which has news reports in media consisting of records of a wide range of socio-economic and political events, viz. ethnic conflicts and human rights violations, over a long period of time, and (b) GTD project \cite{GTDcloud, GTDcodebook}, which has recorded the terrorist attack incidents that occurred in the last half-century across the globe. We aggregate these anti-social events over time, and study the temporal evolution of these events. We present here the results of several analyses like recurrence intervals, Hurst R/S analysis, etc., that reveal the long memory of these time series \cite{Tilak_2012}. Further, we filter the data country-wise, and study the correlations of these anti-social events within the individual countries. Using the multi-dimensional scaling, we cluster the countries together in terms of the co-movements with respect to temporal growths of these anti-social events. The time series of these events reveal interesting statistical regularities and correlations.

The article is organized as follows. Section~\ref{sec_data_method_results} describes the data description, methodology and results in detail. Section~\ref{sec:conclusion} contains the concluding remarks.

%%>>>>>>>>>>>>>>>>>>>>>>>>>>>>>>>>>>>>>>>>>>>>>>>>>>>>>>>>>>>>>

\section{Data description, Methodology and Results}
\label{sec_data_method_results}
\subsection{Data description}
We have used the Global Database of Events, Language, and Tone (GDELT)~\cite{GDELT, GDELTcodebook} which is an open source database hosted and managed by GDELT project through \textit{Google Cloud}. GDELT monitors the world's news media from nearly every corner of every country in print, broadcast, and web formats, in over 100 languages, every moment of every day. The GDELT project is a real-time open database, where the human society is seen through the eyes of the world's news media, reaching deeply into local events, reaction, discourse, and emotions of the most remote corners of the world. The entire GDELT event database is available and can be extracted using \textit{Google BigQuery}. We filtered all events related to ethnic conflicts (EC) and human rights violations (HR) happening around the world spanning over a large time scale. We procured $45,942$ events for EC and $48,295$ for HR for a 15 year period, 2001-2015. 

We have also analysed the data on terrorism events. For the analysis we have used the Global Terrorism Database (GTD) which is an open-source database provides a detailed account of terrorist events around the world from 1970-2017~\cite{GTDcloud, GTDcodebook}. The event database is hosted by the National Consortium for the Study of Terrorism and Responses to Terrorism (START), University of Maryland. We procured $72,521$ events for the same 15 year period, 2001-2015. 

The list of all the countries analysed, containing names along with their three letter ISO codes, is given in Table \ref{table_country}.

%%%================================================
\subsection{Methodology and Results}
\label{sec_methods}

The GDELT and GTD data sets contain detailed information about the anti-social events, viz. ethnic conflicts (EC), human rights violations (HR) and terrorist attacks (GTD), on the scale of a day. Our overarching aim is to observe nature of the memory of each of the time-series (EC, HR and GTD) and the cross-correlations among them, within a country. Further, we would like to group the countries together on the basis of their long-term evolution trends and correlations. First, we study simple statistics of auto-correlations, Hurst R/S analysis and recurrence intervals distribution, of the detrended time series. Later, we study the co-movements of the countries on the events spaces using the multidimensional scaling technique.

\begin{figure}[!h]
   \includegraphics[width=0.49\textwidth]{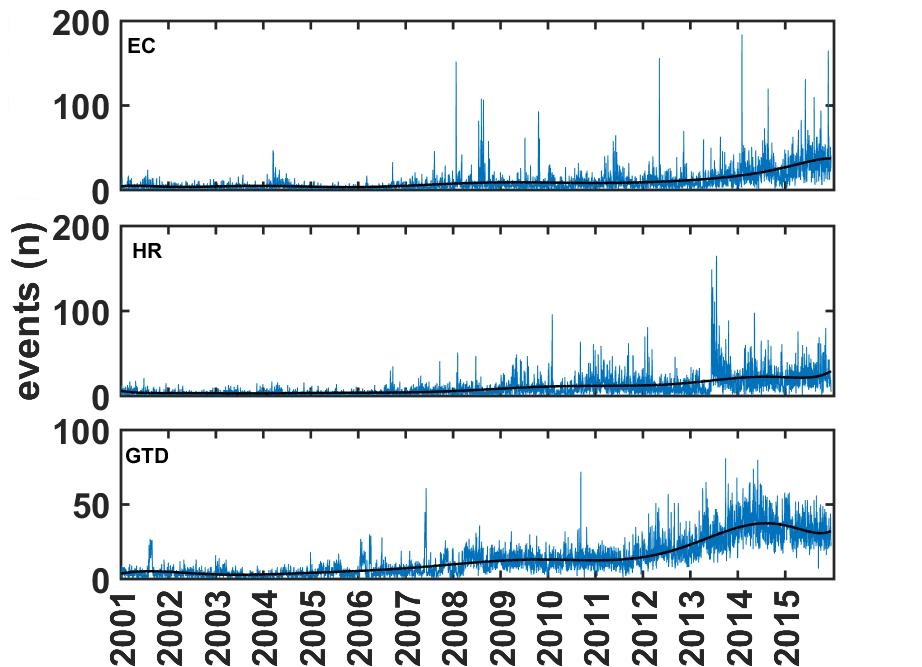}
   \llap{\parbox[b]{2.4in}{(\textbf{a})\\\rule{0ex}{1.6in}}}
   \includegraphics[width=0.47\textwidth]{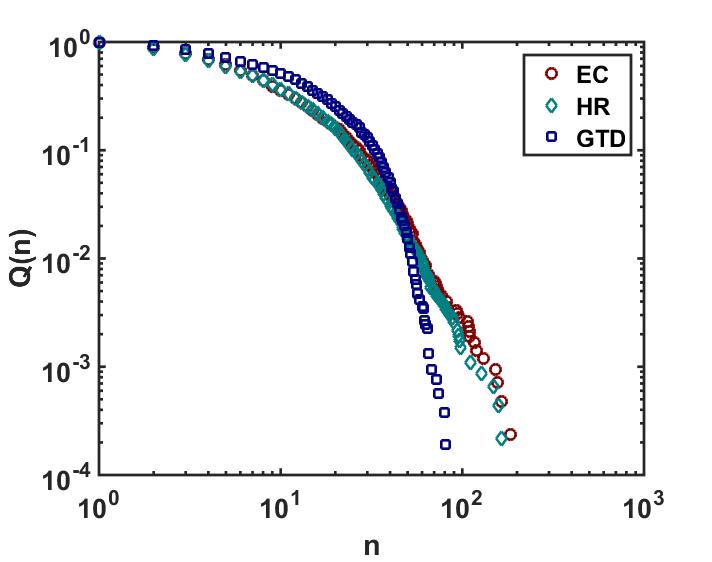}
   \llap{\parbox[b]{2.3in}{(\textbf{b})\\\rule{0ex}{1.6in}}}
\caption{\textbf{(a)} Plots of the time evolution of the number of events $n$ reported daily for EC, HR and GTD during the period 2001-2015 with trends (black solid curves).
\textbf{(b)} The complementary cumulative density function (CCDF) $Q(n$) that $n$ or more events are reported on a particular day. The data seems to fit well to a stretched exponential of the form $(\exp[-an^b])$ for EC (red circles), HR (green diamonds) and GTD (blue squares), with exponents given in Table~\ref{table_CCDF}.}
\label{fig:events}
\end{figure}

We have considered the data for the period 2001-2015, and generated daily time series of EC, HR and GTD,  as shown in Fig.~\ref{fig:events} (a). The black curves show the long time trends, which imply that the time-series are not stationary. To see the spread of the events, we computed the complementary cumulative density function (CCDF) of the events: For the probability density function (PDF) $P(n)$ as $n$ reported events per day, the cumulative density function (CDF) is $F(n) = P(N\leq n)$; then, the CCDF is $Q(n) = 1-F(n)$, such that it estimates the probability of the events are above a particular level $n$, $P(N>n)$. As the empirical PDFs are often too noisy (specially toward the tails) to be relied upon for statistics, it is known that integrating a signal improves its ``Signal-to-Noise ratio''. So, we plot the CCDF as it reduces the noise content and makes the information contained by the signal clearer. 

%%***************************  Events and CCDF *****************************************
%%%********************************************************************

As the number of news entries $n$ per day is a stochastic variable, we often see bursts of activities for all the three anti-social phenomena: EC, HR and GTD. Due to large inter-day fluctuations in the number of reports and the bursty nature, the CCDF shows a broad distribution. Fig.~\ref{fig:events} (b) shows the plots for the CCDF $Q(n$) that $n$ or more events are reported on a particular day, for the three time-series. Each of the curves is well-fitted by a stretched exponential of the form, $(\exp[-an^b])$ with exponents given in Table~\ref{table_CCDF}.

%%********************* CCDF Table***********************************************
\begin{table}[!h]
\centering
\captionsetup{justification=centering}
\caption{Exponent values for events of EC, HR and GTD}
\begin{tabular}{|l|l|l|l|}
\hline
\backslashbox{Series}{Exponents}    & \textbf{a}  & \textbf{b}   \\ \hline\hline
\textbf{EC}  & 0.50$\pm$ 0.01    & 2.49 $\pm$ 0.01   \\ \hline
\textbf{HR}  & 0.76 $\pm$ 0.01   & 2.91 $\pm$ 0.01  \\ \hline
\textbf{GTD} & 0.48 $\pm$ 0.01   & 3.17 $\pm$ 0.01   \\ \hline
\end{tabular}
\label{table_CCDF}
\end{table}

%%********************************************************************

The auto-correlation is the correlation of a signal with a time-delayed copy of itself, as a function of delay or lag. In simple words, it is the similarity between observations as a function of the time lag between them, which can be used for finding repeating patterns or periodicity obscured by noise. 
The Hurst exponent is a popular measure of long-term memory in a time series, which relates to the auto-correlations of the same and the rate at which these auto-correlations decrease as the time-lag between the pair of values increases.

Extreme events are rare in natural as well as social phenomena, but it is essential to study their properties as the consequences of extreme events are often enormous \cite{Santhanam_2008,Chicheportiche_2014,Chicheportiche_2017}. As researchers, we are often interested in the question that how long would we have to wait for extreme events of a certain magnitude to recur. We thus fix a threshold $X^{(q)}$ and consider only the events of magnitude higher than $X^{(q)}$, where $q$ denotes the quantile. 
We define the \emph{recurrence interval} as the time interval between two consecutive extreme events:
\begin{equation}
R_t= \begin{cases}\text{NA}&,X(t)<\Xq\\\inf\left\{\tau>0\mid X(t+\tau)\geq\Xq\right\}&,X(t)\geq\Xq\end{cases},
\label{eq_reccur}
\end{equation}
where $X(t)$ is an event occurring at time $t$, $X^{(q)}$ is the threshold, and $\tau$ is a time lag.

\begin{figure}

 	\includegraphics[width=0.65\textwidth]{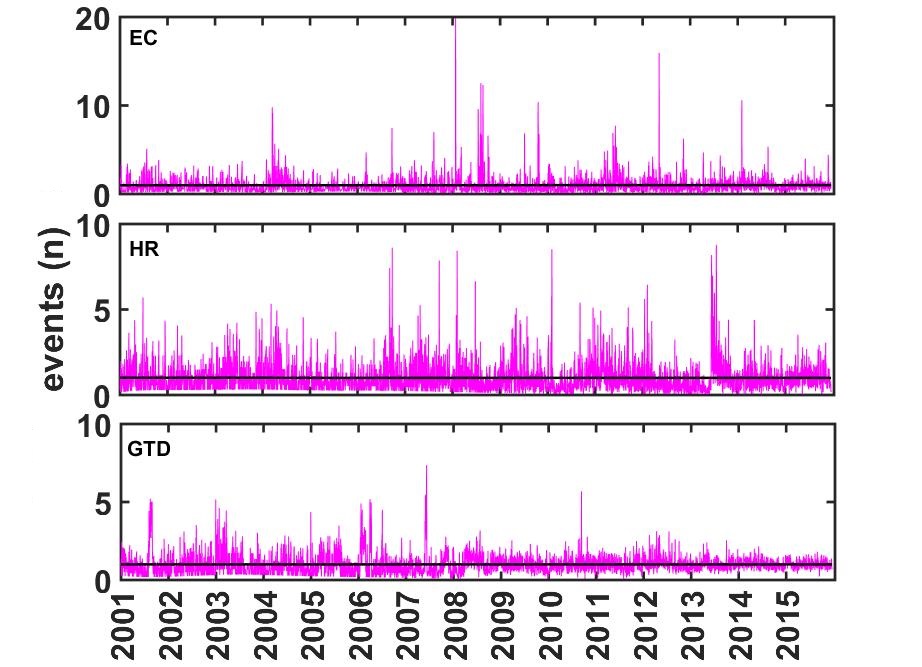}
 	\llap{\parbox[b]{3.1in}{(\textbf{a})\\\rule{0ex}{2.2in}}}
 	\hspace{-0.7cm}
	\includegraphics[width=0.33\textwidth]{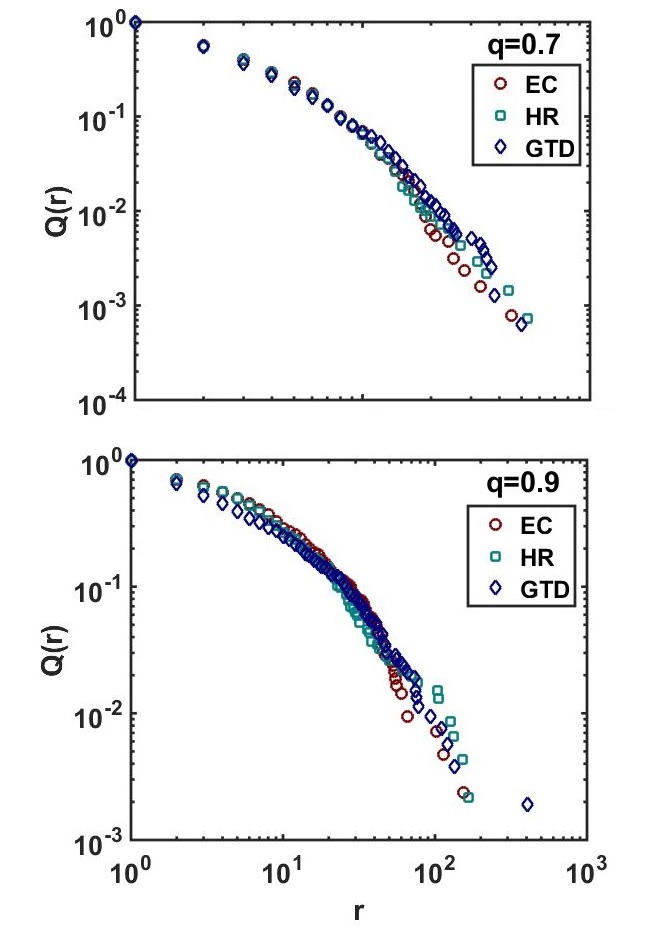}
 	\llap{\parbox[b]{1.6in}{(\textbf{b})\\\rule{0ex}{2.1in}}}
\caption{\textbf{(a)} Detrended time series for EC, HR and GTD with a flat trend (black solid line). \textbf{(b)} shows the plots of complementary cumulative density function  $Q(r)$ that $r$ events recurred at quantiles $q=0.7$ and $q=0.9$. The data for  quantiles $q=0.7$ and $q=0.9$ have been found to fit well to stretched exponentials, $(a\exp[-bn^c])$ with parameters given in Tables~\ref{table_recur07} and \ref{table_recur09}.} 
\label{fig:Reccur}
\end{figure}

A real data series usually exhibit non-stationarity of various forms such as ``seasonal
effects", ``trends", etc. Though it is very difficult to completely eliminate non-stationarity, its effect can be reduced by introducing some corrective measures. Each type of non-stationarity requires a different type of correction. Here, none of the original time series of events is stationary, as can be seen in Fig.~\ref{fig:events}, where the black dashed line shows the inherent trend of the time series. We calculate the trend with a polynomial of degree 10, and then divide the original signal by the computed trend, resulting in a detrended series. Fig.~\ref{fig:Reccur} (a) shows the events time series after detrending it. The black dashed line, which shows trend of this detrended time series, is thus flat.  The detrended events time series can be assumed to be weakly stationary.

%%********************** Recurrence********************************
%%********************************************************************
%%********************************************************************
\begin{table}[!h]
\centering
\captionsetup{justification=centering}
\caption{Parameter values for recurrence CCDF at $q=0.7$}
\begin{tabular}{|l|l|l|l|}
\hline
\backslashbox{Series}{Parameters}    & \textbf{a}  & \textbf{b}  & \textbf{c}  \\ \hline\hline
\textbf{EC}  & 3.53$\pm$ 0.37   & 1.27$\pm$ 0.10  & 0.49 $\pm$ 0.03  \\ \hline
\textbf{HR}  & 3.69 $\pm$ 0.46   & 1.31 $\pm$ 0.12  & 0.49 $\pm$ 0.03   \\ \hline
\textbf{GTD} & 6.95 $\pm$ 0.71   & 1.94 $\pm$ 0.10  & 0.38 $\pm$ 0.01    \\ \hline
\end{tabular}
\label{table_recur07}
\end{table}

%%*******************************************************************

%%********************************************************************
\begin{table}[!h]
\centering
\captionsetup{justification=centering}
\caption{Parameter values for recurrence CCDF at $q=0.9$}
\begin{tabular}{|l|l|l|l|}
\hline
\backslashbox{Series}{Parameters}    & \textbf{a}  & \textbf{b}  & \textbf{c}  \\ \hline\hline
\textbf{EC}  & 1.67$\pm$ 0.09   & 0.54$\pm$ 0.04  & 0.50 $\pm$ 0.02  \\ \hline
\textbf{HR}  & 1.69 $\pm$ 0.10   & 0.56 $\pm$ 0.05  & 0.51 $\pm$ 0.02   \\ \hline
\textbf{GTD} & 4.72 $\pm$ 0.72   & 1.58 $\pm$ 0.15  & 0.28 $\pm$ 0.02    \\ \hline
\end{tabular}
\label{table_recur09}
\end{table}

%%*******************************************************************

We hence analyse the CCDF of the recurrence time intervals on the detrended time series (see Fig.~\ref{fig:Reccur} (a)) to quantify the extreme events or duration of recurrences of an event. 
Fig.~\ref{fig:Reccur} (b) shows the plots of CCDF $Q(r)$ that $r$ events recurred at quantiles $q=0.7$ and $q=0.9$. The data for quantiles $q=0.7$ and $q=0.9$ seem to fit well to stretched exponentials, $(a\exp[-bn^c])$, with parameters given in Tables~\ref{table_recur07} and \ref{table_recur09}. It should be mentioned that as the quantile $q$ increases, the distribution becomes fatter, i.e., lower recurrence time intervals occur less frequently. This observation is quite obviously explained by the fact that at higher
values of $q$ there are fewer extreme events and they are spread apart.

It is often not possible to comprehend certain effects using empirical data. Thus, the results obtained by analyses of empirical data generally need to be compared against standard benchmarks. 
In such situations, artificial data can be simulated according to required specifications and the simulated data can then serve as reliable benchmarks. Therefore, we first use Gaussian noises (white and fractional) to understand certain effects and use them as benchmarks for comparing the empirical statistics.

\begin{figure}
   \includegraphics[width=0.95\linewidth]{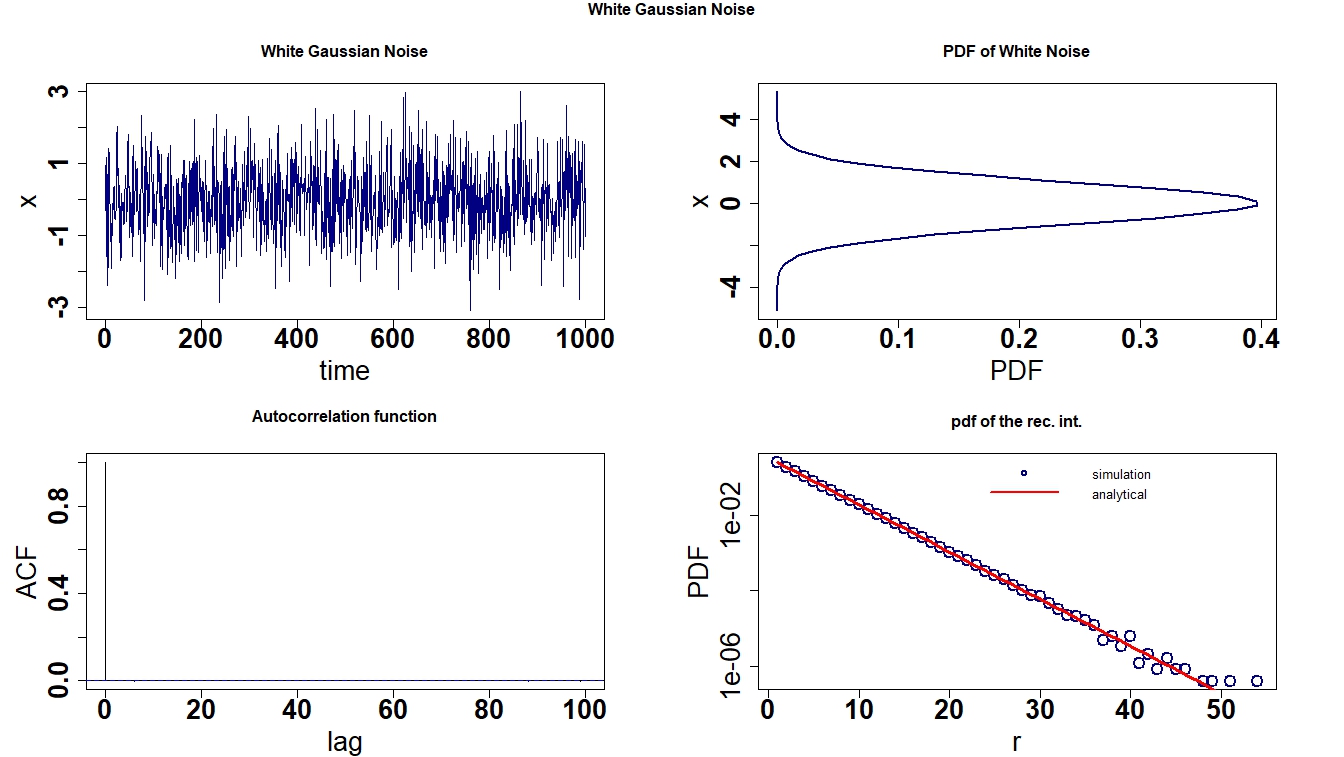}
   \llap{\parbox[b]{4.6in}{(\textbf{a})\\\rule{0ex}{2.2in}}}
   \includegraphics[width=0.95\linewidth]{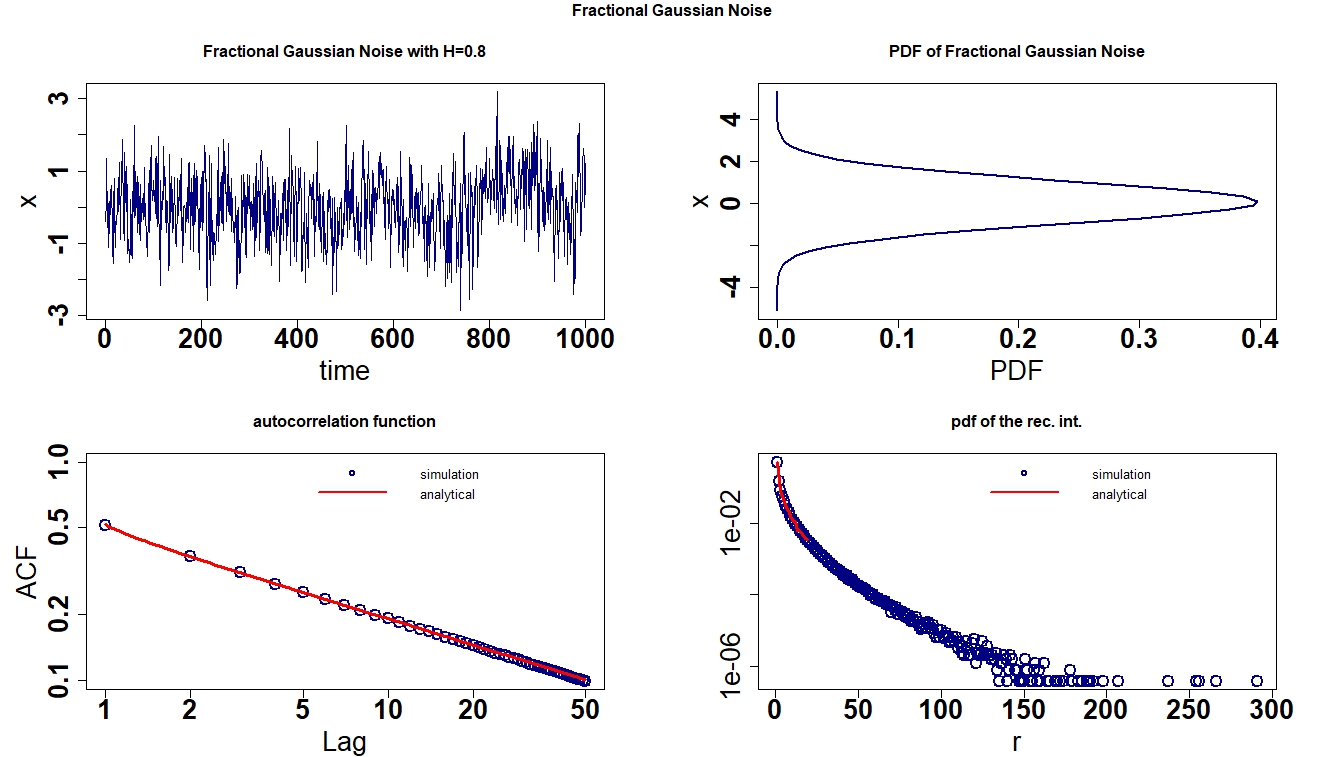}
   \llap{\parbox[b]{4.4in}{(\textbf{b})\\\rule{0ex}{2.2in}}}
\caption{\textbf{(a)} Time series for white Gaussian noise, probability density function for white noise (independent and identically distributed variables), auto-correlation function of time series, probability density function of recurrence time intervals at $q = 0.75$. The time series was generated using the \textit{rnorm()} function in R-software for statistical computing. \textbf{(b)} Time series for fractional Gaussian noise with Hurst index $H=0.8$, probability density function of the fractional Gaussian noise (dependent and identically distributed variables), auto-correlation function of the time series, probability density function for recurrence intervals at $q = 0.75$. The time series was generated using the \textit{simFGN0()} function in R-software for statistical computing. }
\label{fig:benchmark}
\end{figure}

Gaussian noise is a statistical noise having a probability density function equal to that of the Normal (or Gaussian) distribution; a special case is the white Gaussian noise (wGn) or Brownian motion, in which the increments (values at any pair of times) are identically distributed and statistically independent (and hence uncorrelated).
Thus, it has no auto-correlation for positive lags, and an exponentially decreasing recurrence interval distribution. We illustrate a white Gaussian noise in Fig.~\ref{fig:benchmark} (a).

A fractional Brownian motion is a generalization of Brownian motion. The main difference between fractional Brownian motion and regular Brownian motion is that the increments in Brownian motion are independent, whereas increments for fractional Brownian motion are not.
A fractional Gaussian noise (fGn) with Hurst exponent $0 \leq H \leq 1$, is defined as a continuous-time Gaussian process $B_H(t)$ on $[0, T]$, which starts at zero, has expectation zero for all $t$ in $[0, T]$, and has a co-variance function. Mathematically,
\begin{align}
&\forall (t,s)\in \mathbb{R}_+^2, \nonumber\\
&\mathbb{E}[B_H(t)]=0\\
&\mathbb{E}[B_H(t)B_H(s)]=\frac{|t|^{2H}+|s|^{2H}-|t-s|^{2H}}{2}. \label{eq:brownian}
\end{align}
The auto-correlation function (ACF) of a fractional Gaussian noise with Hurst exponent $H$ is given by:
\begin{equation}
ACF(\tau) \rightarrow \dfrac{\mid \tau + 1 \mid ^{2H} + \mid \tau - 1 \mid ^{2H} - 2\mid \tau \mid ^{2H}}{2} .
\label{autocorr_FGN}
\end{equation}
For a stationary process with auto-correlations decaying $ACF(\tau) \sim \tau^{-\gamma}$ (long-memory processes), it can be shown mathematically $\gamma = 2-2H$ \cite{Tarnopolski_2016}.

Fig.~\ref{fig:benchmark} (b) shows the ACF and PDF of recurrence time intervals for fractional Gaussian noise with Hurst exponent $H = 0.8$. If the underlying time series has auto-correlation, then  the extreme events are auto-correlated as well. Evidently the presence of the auto-correlation renders the probability density function of the recurrence time intervals to be a stretched exponential, instead of a pure exponential as observed in the case of white Gaussian noise.  

%%********************************************************************
%%%********************************************************************

The Hurst exponent is a useful statistical method for inferring the properties of a time series.
There are various methods to calculate Hurst exponent, which measures the existence of trend or `persistence' or long-range memory present in the time series. We used the rescaled range (R/S) method to compute the Hurst exponent~\cite{Chakraborti_2005}. 
The rescaled range (R/S) method is calculated for a time series, $X_{1},X_{2},\dots ,X_{T}$, as follows~\cite{torresexponente}: 
We first break the long time series with $T$ data points, into shorter windows of $n$ data points, such that there are $m=T/n$ windows. For each of the $m$ windows of size $n$, we have the partial time series $X_{1},X_{2},\dots ,X_{n}$, for which we calculate the rescaled range:
\begin{enumerate}
\item Calculate the mean $\mu={\frac  {1}{n}}\sum _{{i=1}}^{{n}}X_{i}$
\item Create a mean adjusted series $Y_{i}=X_{{i}}-\mu{\text{  for }}i=1,2,\dots ,n\,$
\item Calculate the cumulative deviate series $Z_{t}=\sum _{{i=1}}^{{t}}Y_{{i}}{\text{  for }}t=1,2,\dots ,n$
\item Compute the range $R(n)=\max \left(Z_{1},Z_{2},\dots ,Z_{n}\right)-\min \left(Z_{1},Z_{2},\dots ,Z_{n}\right)$
\item Compute the standard deviation $S(n)={\sqrt  {{\frac  {1}{n}}\sum _{{i=1}}^{{n}}\left(X_{{i}}-\mu \right)^{{2}}}}$, where $\mu$ is the mean for the partial time series $X_{1},X_{2},\dots ,X_{n}$.
\item Calculate the rescaled range $ R(n)/S(n) $ and average over all the partial time series of length $n$. 
\end{enumerate}
The Hurst exponent is estimated by fitting the power law: $\mathbb{E} [R(n)/S(n)]=Cn^{H}$ to the empirical data, where $C$ is a constant. This can be done by plotting $\log[R(n)/S(n)]$ as a function of $ \log n$, and fitting a straight line. The value of the slope gives the Hurst exponent $H$, such that
 
 \begin{itemize}
\item A value in the range $0\leq H<0.5$ indicates a time series with `anti-persistent' behavior, 
 \item a value in the range $0.5<H\leq 1$ indicates a time series with long-term positive auto-correlation (`persistent' behavior),
\item a value of $H=0$ indicates a pink noise,
\item a value of $H=0.5$ indicates a completely uncorrelated series (Brownian motion).
 \end{itemize}

%%****************ACF **********************************
\begin{figure}[!h]

 	\includegraphics[width=0.95\textwidth]{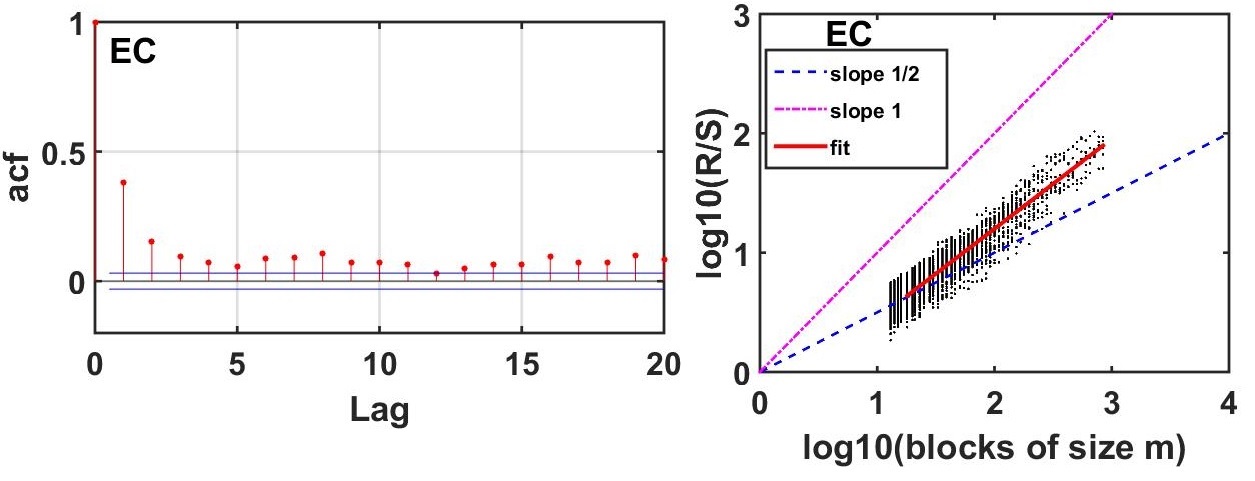}
 	\llap{\parbox[b]{4.6in}{(\textbf{a})\\\rule{0ex}{1.6in}}}\\
  	\includegraphics[width=0.95\textwidth]{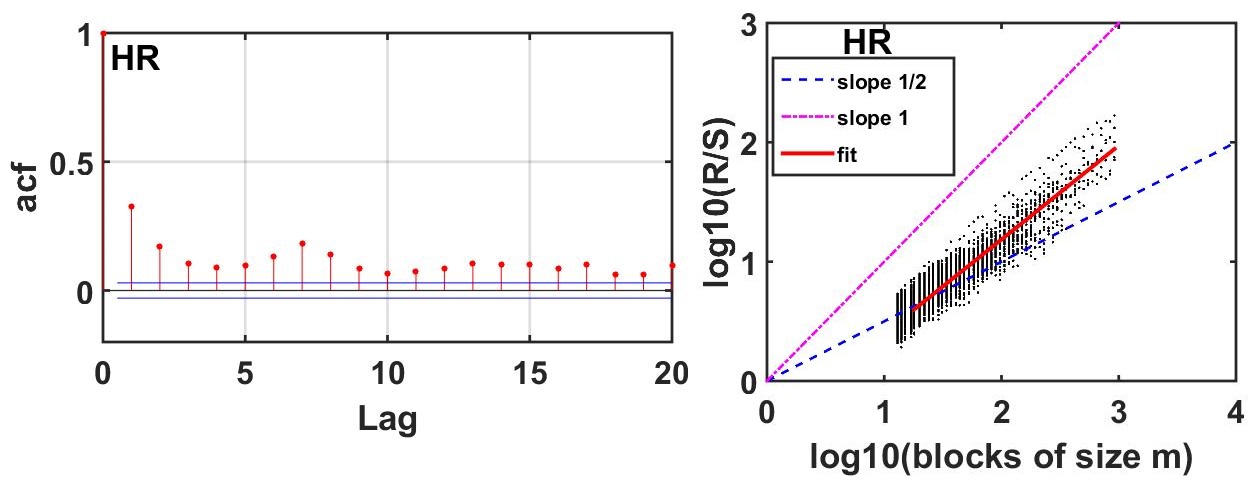}
  	\llap{\parbox[b]{4.6in}{(\textbf{b})\\\rule{0ex}{1.6in}}}\\
 	\includegraphics[width=0.95\textwidth]{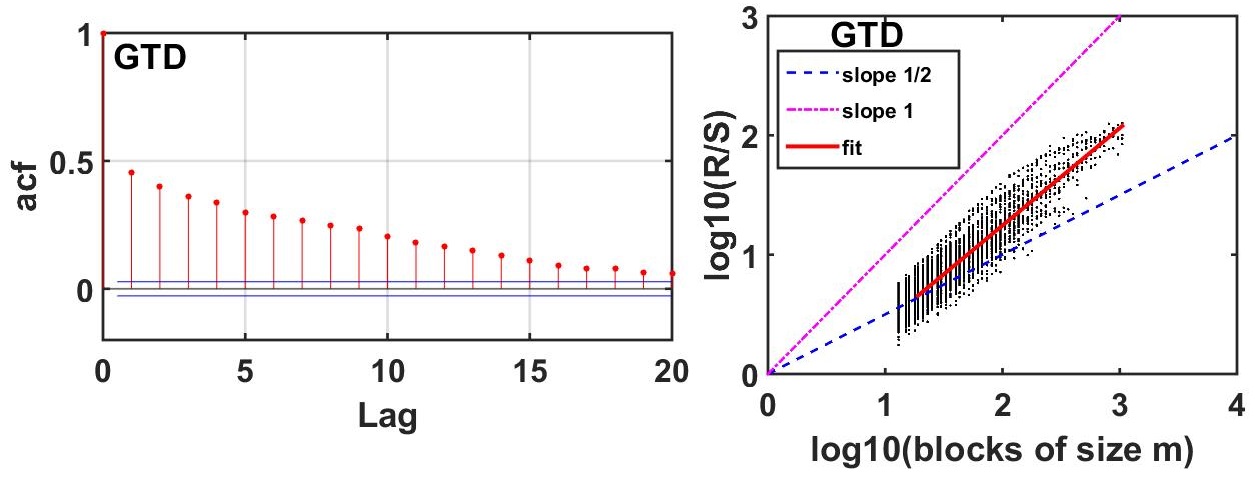}
 	\llap{\parbox[b]{4.6in}{(\textbf{c})\\\rule{0ex}{1.6in}}}
 	\caption{Plot for the auto-correlation of detrended time series and Hurst exponent based on R/S analysis having exponent for \textbf{(a)} EC $0.75 \pm 0.01$, \textbf{(b)} HR $0.78 \pm 0.01$ and \textbf{(c)} GTD $0.82 \pm 0.01$. As the value of the exponent is greater than $0.5$ so the time series shows the persistence behavior for all EC, HR and GTD. }
\label{fig:acf_dfa}
\end{figure}
%%%********************************************************************
Fig.~\ref{fig:acf_dfa} shows the auto-correlation of the detrended time series and Hurst exponent based on R/S analysis having exponent for (a) EC $0.75 \pm 0.01$, (b) HR $0.78 \pm 0.01$ and (c) GTD $0.82 \pm 0.01$.  The auto-correlation for GTD is decaying exponentially. As the value of the exponent is greater than $0.5$ so the time series shows the persistence behavior for all EC, HR and GTD.

\begin{figure}[!h]
 	\includegraphics[width=0.95\linewidth]{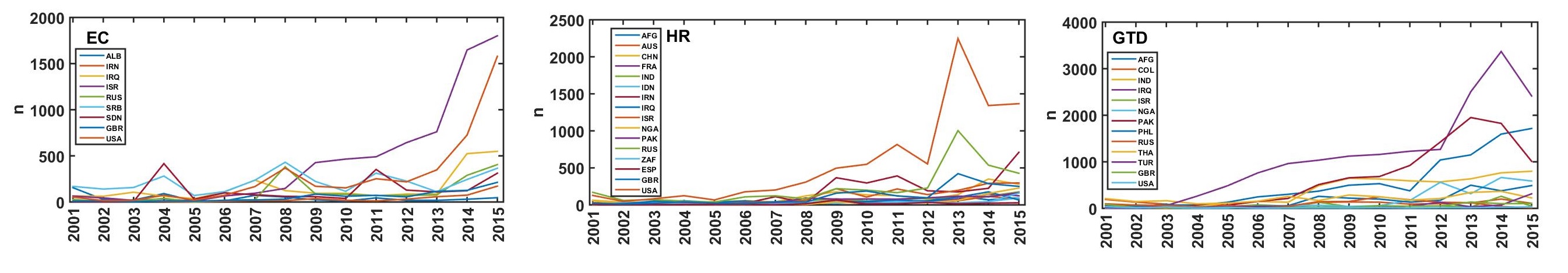}
 	\llap{\parbox[b]{4.7in}{(\textbf{a})\\\rule{0ex}{0.7in}}}
  	\includegraphics[width=0.95\linewidth]{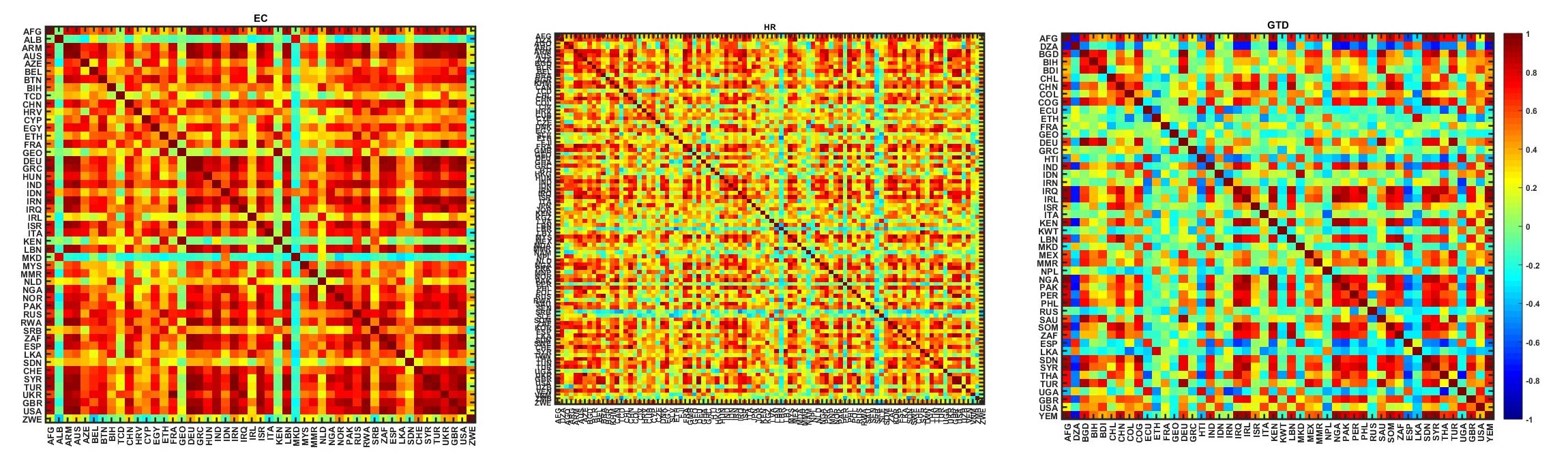}
  	\llap{\parbox[b]{4.7in}{(\textbf{b})\\\rule{0ex}{1.2in}}}
   	\includegraphics[width=0.98\linewidth]{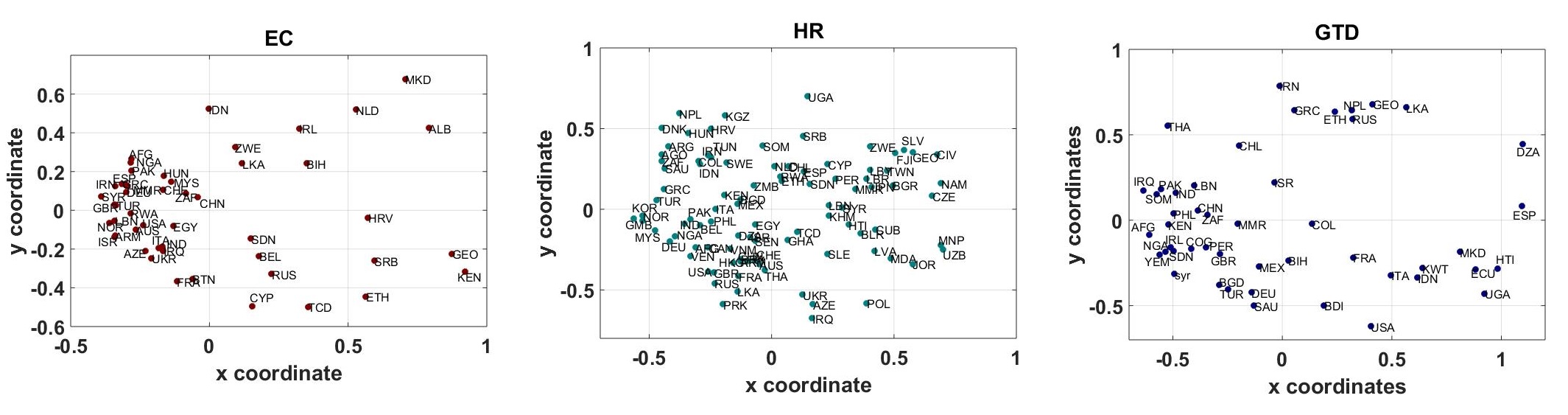}
	\llap{\parbox[b]{4.5in}{(\textbf{c})\\\rule{0ex}{1.2in}}}
\caption{\textbf{(a)} Time series plots for number of events $n$ of different countries for EC, HR and GTD during the period of 2001-15. The list of country names can be seen in Table.~\ref{table_country}. \textbf{(b)} Correlation matrices for EC, HR and GTD. As number of countries $N$ are more than length of time series $T$, i.e. $N>>T$, the power mapping technique with a distortion $\epsilon=0.6$ is applied  to the correlation matrix to suppress the noise. \textbf{(c)} 2D MDS plots for EC, HR and GTD during the period 2001-2015. The MDS plots show the co-movement of the countries: similar countries are grouped together and dissimilar ones placed far apart. }
\label{fig:Corr_MDS}
\end{figure}

Next, we study the co-movements among the different countries across the globe. Fig.~\ref{fig:Corr_MDS} (a) shows the time series of $n$ events (EC, HR and GTD) during the period of 2001-2015 for a few countries chosen arbitrarily. We take $N$ countries and aggregate the events over a year, producing $T=15$ data points for the period $2001-2015$.
To build the correlation matrices, we define the equal-time Pearson cross-correlation coefficient for the time series of the number of events per year $c_i$ as 
\begin{equation}
\rho_{ij}(\tau) = \frac{\langle c_i c_j \rangle - \langle c_i \rangle \langle c_j \rangle}{\sigma_i\sigma_j}.
\label{Eq_corr}
\end{equation}
where $\sigma_i= \sqrt{\langle c_i^2 \rangle -\langle c_i\rangle^2}$ is the standard deviation of $c_i$, $i,j=1, \dots, N$, and $\langle \dots \rangle $ denotes average over the time period $\tau$. The elements $\rho_{ij}$ are restricted to the domain $ -1\leq \rho_{ij}\leq1$, where $\rho_{ij}=1$ signifies perfect correlations, $\rho_{ij}=-1$ perfect anti-correlations, and $\rho_{ij}=0$ corresponds to uncorrelated pairs. 

It is difficult to estimate the exact correlation among $N$  time series, each of length $T$, as spurious correlations or `noise' are present in the finite time series (see Ref.~\cite{Pharasi_2018b}). The quality of the estimation of true correlation in a matrix strongly depends upon the ratio of the length of the time series $T$ and the number of time series $N$, $Q = T/N$. Correlation matrices are less noisy for higher value of $Q $. 
As $N > T$, the corresponding cross-correlation matrices are also singular with $N-T+1$ zero eigenvalues, which leads to poor eigenvalue statistics. Thus, we use the power map technique \cite{Chakraborti_2018, Pharasi_2018a,Pharasi_2018b} to break the degeneracy of eigenvalues at zero and suppress the noise. In this method, a non-linear distortion is given to each cross-correlation coefficient $(\rho_{ij})$ of the  correlation matrix $\boldsymbol \rho$ by: $\rho_{ij} \rightarrow (\mathrm{sign} ~~\rho_{ij}) |\rho_{ij}|^{1+\epsilon}$, where $\epsilon$ is the distortion parameter; here we used $\epsilon=0.6$ (see Refs.~\cite{Pharasi_2018b,Pharasi_2018a} for choice of the parameter).

Fig.~\ref{fig:Corr_MDS} (b)  shows the correlation matrices (after using the power mapping method), computed over different time series across the countries by using Eq.~\ref{Eq_corr}. The correlation matrix for EC shows more correlations (colored red) among the countries as compared to anti-correlations (colored blue). The correlation matrices for HR and GTD look very different. In order 
to visualize the correlations, we apply the multidimensional scaling (MDS) technique. First, we transform the correlation matrix $\boldsymbol \rho$ into distance matrix $\textbf{D}$,  as
\begin{eqnarray}
 d_{ij} = \sqrt{2 (1- \rho_{ij})},
 \label{Eq:distance}
\end{eqnarray}
such that $2 \geq d_{ij} \geq 0$. 
After transforming the correlation matrix into distance matrix, we generate the MDS map.
The MDS algorithm is used to display the structure of similarity in terms of distances, as a geometrical map where each country corresponds to a set of coordinates in
the multidimensional space. MDS arranges different countries in this space according to the strength of the pairwise distances between them. Two similarly behaving countries are represented by two points that are close to each other, and two dissimilarly behaving countries are placed far apart in the map. In general, we choose the embedding dimension to be 2, so that we are able to plot the coordinates in the form of a map. It may be noted that coordinates are not necessarily unique, as we can arbitrarily translate and rotate them, as long as such transformations leave the distances unaffected.
Fig.~\ref{fig:Corr_MDS} (c) shows the 2D MDS plots for EC, HR and GTD based on the similarities/ distances among them. 

%%******************* Corr-MDS **************************************
%%%********************************************************************

At the end, we also calcuted the correlation among the different time series for individual countries. The correlations are computed among EC-HR, EC-GTD and HR-GTD. Few countries like ESP, IDN, ITA and RUS have low correlations among EC-HR, whereas ESP, FRA, ITA, LKA and RUS show anti-correlations for EC-GTD; countries like ESP, FRA, GRC, IDN, ITA, LKA show anti-correlations for HR-GTD. For further details, see  Table.~\ref{table_corr}. It must be noted that these are just linear correlations, and causal relations cannot be inferred.

%%%******************* Correlation Table ******************************
\begin{table}[!h]
\caption{Cross-correlation among events across countries.}
\begin{tabular}{||l||l|l|l|l||l||l|l|l|l|}
\hline 
\textbf{S.No.} & \textbf{Country} & \textbf{EC-HR} & \textbf{EC-GTD} & \textbf{HR-GTD} & \textbf{S.No.} & \textbf{Country} & \textbf{EC-HR} & \textbf{EC-GTD} & \textbf{HR-GTD} \\ \hline \hline
1              & \textbf{AFG }             & 0.72           & 0.80            & 0.84            & 11             &\textbf{ IRQ }             & 0.61           & 0.73            & 0.85            \\ \hline
2              & \textbf{CHN }             & 0.81           & 0.64            & 0.89            & 12             & \textbf{ISR }             & 0.90           & 0.43            & 0.33            \\ \hline
3              & \textbf{DEU}              & 0.65           & 0.95            & 0.49            & 13             & \textbf{ITA }             & 0.21           & -0.05           & -0.09           \\ \hline
4              & \textbf{ESP}              & 0.47           & -0.52           & -0.28           & 14             & \textbf{LKA }             & 0.57           & -0.22           & -0.39           \\ \hline
5              & \textbf{FRA}              & 0.91           & -0.08           & -0.10           & 15             & \textbf{NGA}              & 0.73           & 0.77            & 0.77            \\ \hline
6              & \textbf{GBR}              & 0.81           & 0.72            & 0.84            & 16             & \textbf{PAK}              & 0.83           & 0.65            & 0.82            \\ \hline
7              & \textbf{GRC}              & 0.91           & 0.18            & -0.01           & 17             & \textbf{RUS}              & 0.29           & -0.07           & 0.04            \\ \hline
8              & \textbf{IDN}              & 0.42           & 0.69            & -0.10           & 18             & \textbf{TUR}              & 0.97           & 0.95            & 0.90            \\ \hline
9              & \textbf{IND}              & 0.94           & 0.85            & 0.89            & 19             & \textbf{USA}              & 0.61           & 0.38            & 0.11            \\ \hline
10             & \textbf{IRN}              & 0.80           & 0.11            & 0.45            & 20             & \textbf{ZAF}              & 0.96           & 0.21            & 0.17            \\ \hline
\end{tabular}
\label{table_corr}
\end{table}
%%%********************************************************************
%%%*********************  Country list ***********************************************

\begin{table}[]
\caption{List of countries and their 3-letter ISO codes.}
\begin{tabular}{||l||l|l||l||l|l||l||l|l|}
\hline
\textbf{S.No.} & \textbf{Code} & \textbf{Country}                 & \textbf{S.No.} & \textbf{Code} & \textbf{Country}         & \textbf{S.No.} & \textbf{Code} & \textbf{Country} \\ \hline \hline
1              & \textbf{AFG}  & Afghanistan                      & 36             & \textbf{GEO}  & Georgia                  & 71             & \textbf{NLD}  & Netherlands      \\ \hline
2              & \textbf{AGO}  & Angola                           & 37             & \textbf{GHA}  & Ghana                    & 72             & \textbf{NOR}  & Norway           \\ \hline
3              & \textbf{ALB}  & Albania                          & 38             & \textbf{GMB}  & Gambia                   & 73             & \textbf{NPL}  & Nepal            \\ \hline
4              & \textbf{ARG}  & Argentina                        & 39             & \textbf{GRC}  & Greece                   & 74             & \textbf{PAK}  & Pakistan         \\ \hline
5              & \textbf{ARM}  & Armenia                          & 40             & \textbf{HKG}  & Hong Kong                & 75             & \textbf{PER}  & Peru             \\ \hline
6              & \textbf{AUS}  & Australia                        & 41             & \textbf{HRV}  & Croatia                  & 76             & \textbf{PHL}  & Philippines      \\ \hline
7              & \textbf{AZE}  & Azerbaijan                       & 42             & \textbf{HTI}  & Haiti                    & 77             & \textbf{POL}  & Poland           \\ \hline
8              & \textbf{BDI}  & Burundi                          & 43             & \textbf{HUN}  & Hungary                  & 78             & \textbf{PRK}  & North Korea      \\ \hline
9              & \textbf{BEL}  & Belgium                          & 44             & \textbf{IDN}  & Indonesia                & 79             & \textbf{RUS}  & Russia           \\ \hline
10             & \textbf{BGD}  & Bangladesh                       & 45             & \textbf{IND}  & India                    & 80             & \textbf{RWA}  & Rwanda           \\ \hline
11             & \textbf{BGR}  & Bulgaria                         & 46             & \textbf{IRL}  & Ireland                  & 81             & \textbf{SAU}  & Saudi Arabia     \\ \hline
12             & \textbf{BIH}  & Bosnia-Herzegovina               & 47             & \textbf{IRN}  & Iran                     & 82             & \textbf{SDN}  & Sudan            \\ \hline
13             & \textbf{BLR}  & Belarus                          & 48             & \textbf{IRQ}  & Iraq                     & 83             & \textbf{SEN}  & Senegal          \\ \hline
14             & \textbf{BRA}  & Brazil                           & 49             & \textbf{ISR}  & Israel                   & 84             & \textbf{SLE}  & Sierra Leone     \\ \hline
15             & \textbf{BTN}  & Bhutan                           & 50             & \textbf{ITA}  & Italy                    & 85             & \textbf{SLV}  & El Salvador      \\ \hline
16             & \textbf{CAN}  & Canada                           & 51             & \textbf{JOR}  & Jordan                   & 86             & \textbf{SOM}  & Somalia          \\ \hline
17             & \textbf{CHE}  & Switzerland                      & 52             & \textbf{JPN}  & Japan                    & 87             & \textbf{SRB}  & Serbia           \\ \hline
18             & \textbf{CHL}  & Chile                            & 53             & \textbf{KEN}  & Kenya                    & 88             & \textbf{SWE}  & Sweden           \\ \hline
19             & \textbf{CHN}  & China                            & 54             & \textbf{KGZ}  & Kyrgyzstan               & 89             & \textbf{SYR}  & Syria            \\ \hline
20             & \textbf{CIV}  & Cote D'ivoire                    & 55             & \textbf{KHM}  & Cambodia                 & 90             & \textbf{TCD}  & Chad             \\ \hline
21             & \textbf{COG}  & Democratic Republic of the Congo & 56             & \textbf{KOR}  & South Korea              & 91             & \textbf{THA}  & Thailand         \\ \hline
22             & \textbf{COL}  & Colombia                         & 57             & \textbf{KWT}  & Kuwait                   & 92             & \textbf{TUN}  & Tunisia          \\ \hline
23             & \textbf{CUB}  & Cuba                             & 58             & \textbf{LBN}  & Lebanon                  & 93             & \textbf{TUR}  & Turkey           \\ \hline
24             & \textbf{CYP}  & Cyprus                           & 59             & \textbf{LBR}  & Liberia                  & 94             & \textbf{TWN}  & Taiwan           \\ \hline
25             & \textbf{CZE}  & Czech Republic                   & 60             & \textbf{LBY}  & Libya                    & 95             & \textbf{UGA}  & Uganda           \\ \hline
26             & \textbf{DEU}  & Germany                          & 61             & \textbf{LKA}  & Sri Lanka                & 96             & \textbf{UKR}  & Ukraine          \\ \hline
27             & \textbf{DNK}  & Denmark                          & 62             & \textbf{LVA}  & Latvia                   & 97             & \textbf{USA}  & United States    \\ \hline
28             & \textbf{DZA}  & Algeria                          & 63             & \textbf{MDA}  & Moldova                  & 98             & \textbf{UZB}  & Uzbekistan       \\ \hline
29             & \textbf{ECU}  & Ecuador                          & 64             & \textbf{MEX}  & Mexico                   & 99             & \textbf{VEN}  & Venezuela        \\ \hline
30             & \textbf{EGY}  & Egypt                            & 65             & \textbf{MKD}  & Macedonia                & 100            & \textbf{VNM}  & Vietnam         \\ \hline
31             & \textbf{ESP}  & Spain                            & 66             & \textbf{MMR}  & Myanmar                  & 101            & \textbf{YEM}  & Yemen            \\ \hline
32             & \textbf{ETH}  & Ethiopia                         & 67             & \textbf{MNP}  & Northern Mariana Islands & 102            & \textbf{ZAF}  & South Africa     \\ \hline
33             & \textbf{FJI}  & Fiji                             & 68             & \textbf{MYS}  & Malaysia                 & 103            & \textbf{ZMB}  & Zambia           \\ \hline
34             & \textbf{FRA}  & France                           & 69             & \textbf{NAM}  & Namibia                  & 104            & \textbf{ZWE}  & Zimbabwe         \\ \hline
35             & \textbf{GBR}  & United Kingdom                   & 70             & \textbf{NGA}  & Nigeria                  &                & \textbf{}     &                  \\ \hline
\end{tabular}
\label{table_country}
\end{table}
%%%**************************************************************************
%%>>>>>>>>>>>>>>>>>>>>>>>>>>>>>>>>>>>>>>>>>>>>>>>>>>>>>>>>>>>>>
\section{Concluding remarks}
\label{sec:conclusion}
%Reports in news media serve as fair proxy for the importance and intensity of events, from the number of reports and lingering span of time through which the reports follow. 
In this paper, our goal was to do the time series analysis and apply data science approaches to the study of the daily anti-social events like ethnic conflicts (EC), human right violations (HR) and terrosrist attacks (GTD). As the time series were non-stationary, so we made them stationary by detrending them. We computed the recurrence interval distribution of events and made attempts to relate it with its auto-correlation function. Then we computed the Hurst exponent using the rescaled range (R/S) analyses, which gives the information about whether long memory is present or not. Further, our interest was to study the co-movements of the countries in the respective events spaces. To visualize the co-movements, we computed the cross-correlations among different countries, transformed the correlations  into distances and then projected the distances into 2D multidimensional scaling maps.  
%%>>>>>>>>>>>>>>>>>>>>>>>>>>>>>>>>>>>>>>>>>>>>>>>>>>>>>>>>>>>>>
\begin{acknowledgement}
The authors would like to thank Anirban Chakraborti, Vishwas Kukreti, Arun S. Patel and Hirdesh K. Pharasi for critical discussions and inputs. 
KS acknowledges the University Grants Commission (Ministry of Human Resource Development, Govt. of India) for her senior research fellowship. 
SSH and KS acknowledge the support by University of Potential Excellence-II grant (Project ID-47) of JNU, New Delhi, and the DST-PURSE grant given to JNU by the Department of Science and Technology, Government of India.
\end{acknowledgement}
%%>>>>>>>>>>>>>>>>>>>>>>>>>>>>>>>>>>>>>>>>>>>>>>>>>>>>>>>>>>>>>

\bibliographystyle{spmpsci}
\bibliography{main}

\end{document}